# Multi-source and Multimodal data fusion for predicting academic performance in blended learning university courses

Wilson Chango[a], Rebeca Cerezo[b], Cristóbal Romero[c*],
[a] Pontifical Catholic University of Ecuador, Department of Systems and Computing,
[b] University of Oviedo - Spain, Department of Psychology,
[c] University of Córdoba – Spain, Department of Computer Science,  *cromero@uco.es

**ABSTRACT**

In this paper we applied data fusion approaches for predicting the final academic performance of university students using multiple-source, multimodal data from blended learning environments. We collected and preprocessed data about first-year university students from different sources: theory classes, practical sessions, on-line Moodle sessions, and a final exam. Our objective was to discover which data fusion approach produced the best results using our data. We carried out experiments by applying four different data fusion approaches and six classification algorithms. The results showed that the best predictions were produced using ensembles and selecting the best attributes approach with discretized data. The best prediction models showed us that the level of attention in theory classes, scores in Moodle quizzes, and the level of activity in Moodle forums were the best set of attributes for predicting students' final performance in our courses.

## 1  Introduction

Blended learning (b-learning) is a method of teaching approach that combines online learning with traditional in face-to-face classroom methods. In research literature [1] the terms blended learning, hybrid learning and mixed-mode instruction are often used to refer b-learning. Its main goal is to overcome the drawbacks of pure online learning and it remains a priority issue in technology enhaced education despite having been put into practice on all over the word for 20 years ago. Nowadays, in the current pandemic scenario, blended instruction has become more important since it is going to be the new normal in terms of teaching-learning in higher education. COVID-19 has led to the sudden suspension of teaching activities in many countries and the scramble to find new ways to resume classes with restrictive space and hygiene requirements. Many institutions do not have enough space to implement the necessary measures and are forced to rethink their face-to-face learning plans as blended learning plans.

Rapid advances in technology have let us capture all student actions in their interactions with virtual and traditional learning environments. Blended learning environments gather a huge amount of data about students' multimodal interactions in traditional classrooms and on-line environments from a wide range of data sources [2]. So, these data sources need to be fused and mined to shed light on educational issues such as prediction of student performance. In this line, Educational data mining (EDM) [3] can be applied to discover and improve educational processes from information extracted from educational data, which is then used to understand the educational process [4]. EDM has been widely used to improve and enhance learning quality, as well as in the pursuit of research objective to understand the teaching-learning process [5]. In this line, one of the most frequent and the oldest studied tasks/problems in EDM is the prediction of learners' performance. It is still a challenge to predict student learning achievement in blended learning environments combining online and offline learning [6] making data fusion techniques necessary.

In this study we do a classification task for predicting the value of a categorical/nominal attribute (the class or final academic status of the student: Pass, Fail or Drop out) based on other attributes (the predictive attributes from various available data sources). We propose applying different data fusion approaches and classification algorithms to data gathered from several sources (theory classes, practical sessions, online sessions and final exams) in a blended university course in order to predict the students' final academic performance. The research questions posed by this study are:

Question 1.- Which data fusion approach and classification algorithms produce the best results from our data?

Question 2.- How useful are the prediction models we produce to help teachers detect students who are at risk of drop out or fail the course?

This paper is organized as follows. The first section covers the background of the related research areas. Following that, we describe the proposed methodology, and describe the data used and how it was preprocessed. Next, we describe the experiments we performed and the results they produced. Finally, we discuss the implications, conclusions and future research.

## 2  Background

Multimodal Learning Analytics (MLA) is a subfield of Learning Analytics (LA) that uses data from different sources about learning traces for doing a single analysis. MLA is much related with multi-view, multi-relational data, and data fusion. It is used for understanding and optimizing teaching-learning process in which the use of videos has now been consolidated, from traditional courses to mixed and online courses [7]. It has become increasingly broadly applied in both online and in face-to-face learning environments where interactions are not solely mediated by digital devices [8]. MLA uses log-files and gaze data, biosensors, interactions with videos, audio and digital documents, and any other data source for understanding or measuring the learning process. So, one important issue to resolve is how to combine, or fuse, the data extracted from several sources/modalities in order to provide a better and more comprehensive view of teaching-learning processes [9].

Data/Information fusion is the process of efficiently transforming and integrating information gathered from various sources at various times, either automatically or semi-automatically, into a form that can provide practical support to a decision-making process, be that human or automatic. Data fusion is used for reducing the dimension of size of the data, optimizing how much data/info there is, and extracting information that is useful [10]. Multimodal data fusion is the combination/integration of data from different/several sources/modalities/contexts in order to obtaining a better understanding of the teaching-learning process [11]. There are three main types of multimodal fusion approaches [12]:

- **Naïve fusion** is the simplest approach. It builds several classifiers using features summary statistics obtained from each of the different data sources/streams.
- **Low-level fusion frame (or feature fusion)** merges raw data. It synchronizes the data sources/streams at each time stage and it analyses the features after their integration together.
- **High-level fusion frame (or quasi feature-level)** uses semantic analysis first to attempt to make sense of the raw data. It extracts one or several abstract or high level features starting from one or more data sources/streams before integrating them.

A different way to group/classify fusion methods is by considering the fusion periods/steps. In terms of period-level fusions, to date there are the next subtypes of MLA fusion [13]:

- **Feature-level or early fusion.** This happens at the first steps of multimodal data fusion in which there is concatenation and no overlapping. So, the obtained feature/attribute vectors are heterogeneous due to concatenate different data sources/streams.
- **Decision-level or later fusion.** In contrast to feature-level fusion, decision-level fusion is conducted at a later step. It allows each data source/stream to use the most appropriate classifier for its features. The drawback is that it can be hard and time consuming to fuse different classifiers at this step.
- **Hybrid fusion**. This type of fusion propose to use in a hybrid way both feature-level and decision-level fusions.

Finally, most data fusion schemes have four stages; preprocessing the data set, shrinking the dimensions of the data and using data correlation to identify the most fruitful feature sets, training classification algorithms, and finally forecasting new data based on classification algorithms. Feature selection algorithms are normally used in data fusion for classification problems in order to reduce dimensions of data and to produce the best results [14].

In this study we apply several multimodal fusion approaches based on Naïve and decision-level fusion. To our knowledge, there are no studies examining how data fusion approaches can help for predicting students' performance in blended learning.

## 3 Proposal

This paper proposes to use a data fusion and mining methodology for predicting students' final performance starting from multi-source and multimodal data (see Figure 1).

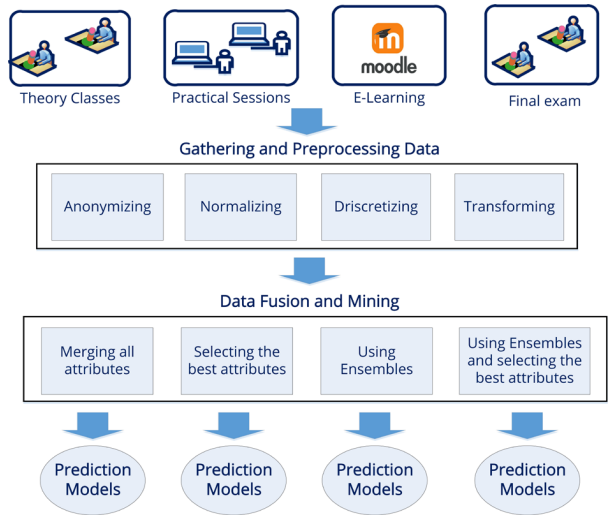

**Figure 1. Proposed data fusion and mining methodology for predicting students' performance from multiple data sources.**

There are two main stages in our methodology as we can see Figure 1.

- First stage: It gathers data from several sources: theory classes, practical sessions, online sessions with Moodle, and the course final exam. It also applies some pre-process tasks (anonymization; attribute normalization and discretization; and format transformation) for generating datasets in two formats: numerical and categorical.

- Second stage: It uses different data fusion approaches (merging all attributes; selecting the best attributes; using ensembles; and using ensembles and selecting the best attributes) and several white-box classification algorithms with the datasets. Then, we compare the predictions produced by the models in order to discover the best approach and classification model so that it is used for predicting students' final performance.

## 4  Data

We used information from 57 electrical engineering first-year students from University of Cordoba (UCO-Spain) in the Introduction to Computer Science course during the first semester of academic year 2017-2018. The main contents of the course were: History of Computer Science; Introduction to Operating Systems, Databases, Internet and Office Applications; and Introduction to Programming.

## 4.1 Gathering Data

We have gathered all the information from four data sources of the same course: theory classes, practical classes, on-line sessions and final exam. The first three data sources gave us the input attributes and the final exam, the output attribute or class to predict. In this course there was only one group for theory classes and two groups for practical classes. A single teacher collected all the data and video recorded the theory classes because the same teacher was assigned to all the groups in this course. The students all gave their written consent to being recorded, after being informed about the study, and to have their data from practical and online sessions in Moodle collected for the study.

### 4.1.1 Theory classes

Theory classes are the traditional face-to-face sessions in which the teacher teaches the theoretical content of the course using blackboards/whiteboards/projectors.

We collected the following information by extracting it from videos (see Figure 2) recorded during the 15 in person theory classes given by the course teacher:

- Theory.Attendance: This was gathered manually from the videos. The value was 1 for a student attending a session, and 0 for a student not attending a session.

- Theory.Location: This was gathered manually from the videos. This value was which row the student sat in (from 1 to 12 rows of chairs or 0 if the student did not attend) in the classroom.

- Theory.Attention: This was gathered semi-automatically from the videos. This value measured how much time the student spent looking at the instructor on each theory class (out of 110 minutes of each lesson).

- Theory.TakeNotes: This was gathered semi-automatically from the videos. This value measured how much time the student spent typing notes or writing during each theory class (out of 110 minutes of each lesson).

The teacher recorded all of the theory classes using a camera on the lecturer's desk (see Figure 2). We also used the individual photos of each student provided in Moodle in order to recognize them. Two researchers involved in the study analyzed the 1650 minutes of recorded video to ensure reliability.

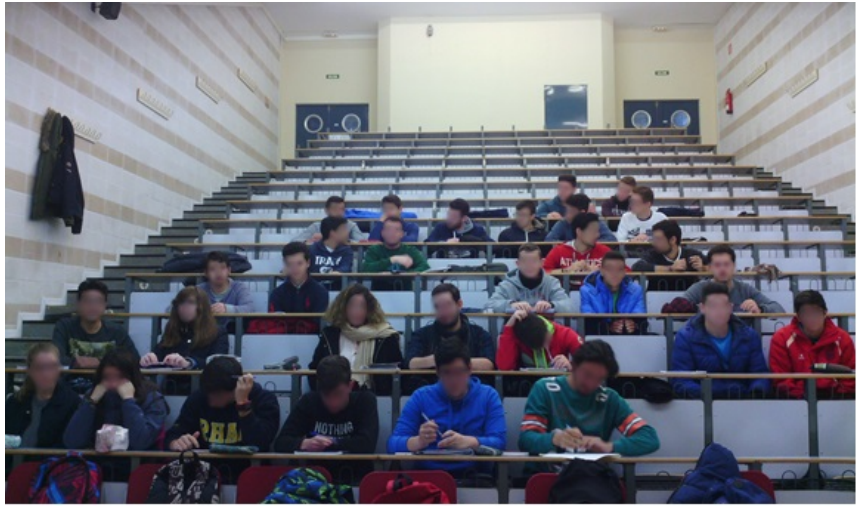

**Figure 2. A snapshot of a theory classroom**

We created two specific programs in Python to semi-automatically produce the attention and note-taking variables. The first program detected the proportion of time a student's face was facing forwards. The second program detected the proportion of time a student's pen was vertical. It was not possible to detect these for all of the students simultaneously, so these two programs were executed for specific coordinates for each student's head and hands (57 executions per program). Although the time values produced were not one hundred percent accurate, they were very close to what we observed. When looking at the videos we noted that there were times when students had their eyes closed, or were looking forwards but not at the teacher or the blackboard or slides, and occasionally students had their pens in their hands without writing. An Excel file was produced with the values of each attribute for each student in each theory class session.

### 4.1.2 Practical sessions

Practical sessions were those in which the students applied their theoretical learning, such as using two operating systems (Windows and Linux), two office applications (Excel and Access) and a visual programming interface (IDE in Python). We selected Python due to it is a high-level and general-purpose programming language for Engineers. In fact, it is one of the current most popular programming languages thanks to the language versatility, scripting & automation, and it is simple and easy to learn

The own professor of practices collected the following information about the 10 face-to-face practical sessions:

- Practical.Attendance: This was obtained starting from the signing sheet used to monitor each day's attendance. In this course, there were 5 practical subjects spread over 10 hours of lessons on 10 face-to-face practical sessions. The value was 1 for a session the student attended, and 0 for a session the student did not attend.
- Practical.Score: This was each students' score from each practical subject, graded by the teacher for each of the five practical subjects. The values were between 1 and 10.

The teacher provided us an Excel file with student attendance and scores.

### 4.1.3 Online sessions

Students also interacted with Moodle for accessing all of the complementary online resources provided by the teacher, including slide-show files for each theory class, a description of each practical, forum discussions, online activities, and quizzes.

The following information was obtained from Moodle logs [15] about student interaction with the online course:

- Moodle.Quiz: This was the students' scores obtained in a Moodle multiple-choice test set by the teacher to test each students' performance in the middle of the course. This was a value between 0 and 10.
- Moodle.Forum: This was the number of contributions/actions each student made to the Moodle discussion forum for the course, either consulting their peers, asking, or answering questions. This ranged from 0 to a maximum value provided by the most active student in the forum.
- Moodle.Task: This was the number of activities that each student uploaded into the Moodle. The instructor requested the students to complete 5 compulsory and 3 optional activities. This variable ranged from 0 to 8.
- Moodle.Time: This was the total time that each student spent logged/connected in Moodle. Each time that a student login to Moodle began a new work session, and the connection time was recorded. This value ranged from 0 to the time spent by the student who spent most time connected to the platform. In some cases user do not explicitly close the session but instead directly close the browser window which produces false values. We solved this kind of problem with outliers in our data files by using specific pre-processing algorithms [16].

The teacher downloaded the log file of the course from the Moodle interface and we automatically gathered the values for each student by using a specific tool for preprocessing Moodle logs that we had developed for a previous study [17]. This tool generated an Excel file with these four attributes for each student who accessed Moodle.

#### 4.1.4 Final exam

The final exam is the in situ final examination that the students had to do at the end of the course. The exam had two parts: a theory part, on paper, with 6 questions (3 multiple choice, 3 open answer) in one hour; and a computer-based practical part, requiring the students to solve 4 problems in 1 hour. The final score from the exam was the sum of the scores in each part, which was given as a score out of 10.

The teacher provided an Excel file with the students' marks in the final exam.

## 4.2 Preprocessing Data

We preprocessed [18] all of the data in the aforementioned Excel files. Firstly, the data were anonymized, We implemented a basic solution, using a randomly generated number as a user Identification (ID) rather than the users' names, and replaced the students' names with the random ID in the four Excel files.

Then the input attributes were normalized/rescaled. In this case we rescaled/normalized all of the input attribute values to the same range [0-1] by using the well-known Min-Max method, which is a linear conversion of the data using the formula: $Zi = Xi - \frac{min(X)}{max(X)} - min(X)$, where $X = (x1,...,xn)$.

Next, the output attributes and input attributes were discretized. We stored the 10 input attributes both in numerical and categorical formats. In order to do discretization we used the well-known Equal-Width binning with the following 3 bins/labels: *High*, *Medium* and *Low*. This method divides all the possible values into only *N* subranges of the same size using the equation: $bin_{width} = \frac{Max\ Value - Min\ Value}{N}$.

We also discretized the output attribute or class to predict (the students' final academic performance or status). We used a manual method in which the own user/instructor had to specify the cut-off points. In our case, the class had the following 3 values and cut-off points:

- *PASS*: Students scoring 5 or more out of 10 in the final test. In our case, this was 19 out of 57 students (33.33%).
- *FAIL*: Students scoring less than 5 out of 10 in the final test. In our case, this was 17 out of 57 students (29.82%).
- *DROPOUT*: Non-completing students who chose not to do the compulsory final test, and thus did not successfully complete the course [19]. In our case, this was 21 out of 57 students (36.84%).

Finally, we converted the files from Excel to CSV (Comma-separated values) files. It is a delimited text file in which each line of the file is a full data record and it uses a comma character/symbol in order to separate values. We transformed each of the two versions of the four Excel files (numerical and categorical values) into CSV files because they can be directly opened and used by the Weka data mining framework that we used in the experiments [20].

## 5 Experiments and Discussion

We carried out four different experiments using four data fusion approaches and several classification algorithms with the preprocessed numerical and discretized data to predict academic performance in a university course (see Figure 3).

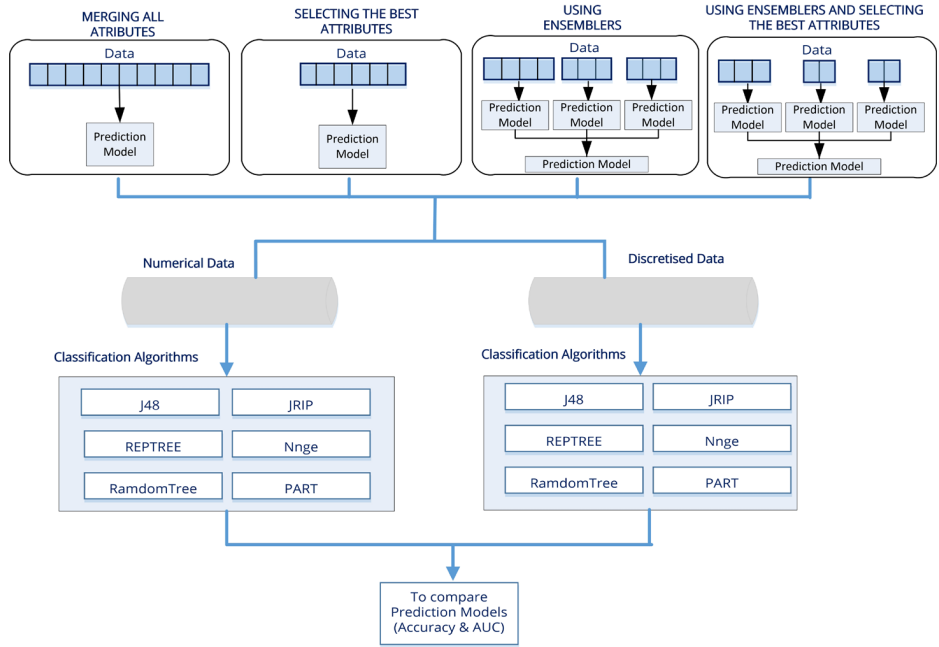

**Figure 3. Visual description of the experiments**

We used two types of white box classification models: rule induction algorithms and decision trees. The obtained models by these algorithms (IF-THEN rules de Decision Trees) are simple and clear and so, they are easy to understand by humans. On the one hand, IF-THEN classification rules provide a high-level knowledge representation that is used for decision making. On the other hand, decision trees can also be converted into a set of IF-THEN classification rules. In our experiments, we selected six well-known classification algorithms integrated in Weka data mining tool [20]: three decision trees algorithms (J48, REPTree and RandomTree) and three rule induction algorithms (JRip, Nnge and PART).

We executed each algorithm using a k-fold cross-validation (k=10). In this way, the dataset is randomly divided into 10 disjointed subsets of equal size in a stratified manner. Of the k (10) subsamples, a single subsample is used as the validation data for testing the model and the remaining other k-1 (9) subsets are used/combined to form the training data. This process is repeated k (10) times and the result is averaged in one single estimation. In order to compare the prediction performance of the algorithms, we needed to select some specific classification metrics from all those previously used in the literature [21]. Some of the most popular evaluation metrics for classification are: Accuracy(Acc), Precision, Recall, Specificity, F-measure, F1-score, Log Loss, Geometry mean, Area Under the Curve(AUC), etc. We selected the following:

- **Accuracy(ACC)** is the most used traditional method to evaluate classification algorithms. It provides a single-number summary of performance. In our case, it is obtained by the next equation: $Acc=\frac{\text{Number of students correctly classified}}{\text{Total number of students}}$. This metric show the percentage of correctly classified students.
- **Area under the ROC curve(AUC)** measures the two dimensional area underneath the entire Relative Operating Characteristic (ROC) curve. ROC curve lets to find possibly optimal models and to discard the suboptimal ones. AUC is often used when the goal of classification is to obtain a ranking because ROC curve construction requires to produce a ranking.

## 5.1 Experiment 1: Merging all attributes

In experiment 1 we applied the classification algorithms to a single file with all the attributes merged. Firstly, we fused the different values (for each session) of the 6 attributes collected in the theory and practical sessions in order to have just one single value for each attribute. In our case, we had 15 values (15 lectures) for each one of the 4 attributes collected in the face-to-face theory classes and 10 (10 sessions) and 5 (5 practicals) values respectively for each of the 2 attributes for face-to-face practice sessions. Fusing the 4 4 values about the on-line sessions was not necessary because the specific tool that we used for preprocessing the Moodle logs [17] gave a single value for each attribute directly. To fuse the numerical values, we averaged, that is, we calculated the arithmetic mean by summing of all of the values and dividing by the number of values. In order to merge the discrete or categorical values, we used the mode; the value that appeared most often. It was not necessary to do anything to the files containing the students' academic performance or course status. Following that, we merged the four CSV files into a single CSV file by combining the fused values from each row with the same ID number (without adding the ID number itself) for each file. The same procedure was used for the numerical and discrete/categorical CSV files in order to generate two summary datasets. Each dataset has ten input attributes (in numerical or discrete format) and only one output attribute or class. Finally, we executed six classification algorithms on the two summary datasets and we produced the results (%Accuracy and ROC Area) shown in Table 1.

**Table 1. Results produced by merging all attributes.**

|  | NUMERICAL DATA | | DISCRETIZED DATA | |
|---|---|---|---|---|
|  | % Accuracy | AUC | %Accuracy | AUC |
| **Jrip** | 77.1930 | 0.8440 | 78.9474 | 0.8880 |
| **Nnge** | **80.4561** | 0.8760 | 75.4386 | 0.8630 |
| **PART** | 78.9474 | 0.8640 | **80.4561** | **0.9170** |
| **J48** | 75.4386 | 0.8640 | 78.9474 | 0.8780 |
| **REPTree** | 75.4386 | 0.8630 | 76.6667 | 0.8480 |
| **Randomtree** | 70.1754 | 0.7820 | 73.6842 | 0.8180 |
| **Avg.** | 76.2748 | 0.8488 | 77.3567 | 0.8686 |

Table 1 shows that the best results (highest values) were produced by Nnge (80.45 %Acc) and Part (80.45%Acc and 0.91 AUC) algorithms. On average, most of the algorithms exhibited slightly improved performance in both measures when using discretized data.

## 5.2 Experiment 2: Selecting the best attributes

In experiment 2 we applied the classification algorithms to a single file with only the best attributes.

Firstly, we applied attribute selection algorithms to the summary files from experiment 1 in order to eliminate redundant or irrelevant attributes. That helps to find the optimal feature set most strongly correlated with the class to predict. The selection of characteristics is important in the classification process by reducing not only the dimensions of the characteristic set but also the additional calculation time required by the classification algorithms. We used the well-known *CfsSubsetEval* (Correlation-based Featured Selection) method provided by the Weka tool [20]. This assesses the merit of attribute subsets by looking at the predictive capacity of each feature in the subset and how redundant they are. In this way, it selects the features that are more correlated with the class. Starting from our initial 10 input attributes, we produced two sets of 3 different optimal attributes (see Table 2) for the numerical and discretized datasets.

**Table 2. Results of the attribute selection with CFSSubsetEval.**

| Dataset | # selected features | Name of Selected features |
|---|---|---|
| Normalized | 3 | Theory.Location<br>Moodle.Quiz<br>Theory.Notes |
| Discretized | 3 | Theory.Attention<br>Moodle.Quiz<br>Moodle.Forum |

Following that, we executed the six classification algorithms with the two new summary datasets producing the results (%Accuracy and ROC Area) shown in Table 3.

**Table 3. Results obtained when selecting the best attributes.**

|  | NUMERICAL DATA | | DISCRETIZED DATA | |
|---|---|---|---|---|
|  | % Accuracy | AUC | %Accuracy | AUC |
| **Jrip** | 80.7018 | 0.8490 | **82.4561** | 0.9140 |
| **Nnge** | **82.4561** | 0.9140 | 78.9474 | 0.8430 |
| **PART** | 77.1930 | 0.8750 | 80.7018 | 0.9140 |
| **J48** | 80.7018 | 0.8680 | **82.4561** | **0.9230** |
| **REPTree** | 77.1930 | 0.8940 | 78.9474 | 0.8880 |
| **Randomtree** | 75.4386 | 0.8320 | 82.4561 | 0.9170 |
| **Avg.** | 78.9473 | 0.8720 | 80.9941 | 0.8998 |

Table 3 shows that the best results (highest values) were produced by Jrip (82.45%Acc), Nnge (80.45 %Acc), and J48 (82.45 %Acc and 0.92 AUC) algorithms. Again, on average most of the algorithms exhibited slightly improved performance in both measures when using discretized data.

## 5.3 Experiment 3: Using ensembles

In experiment 3 we applied an ensemble of classification algorithms to each different source of data. First, we created three different sets of datasets starting from the fused attribute values in experiment 1. However, instead of merging all of the attributes from the 4 data sources into a single file, we added the students' final academic status to each dataset. This produced three sets of datasets (6 files in total): two files (numerical and discrete version) for the theory classes with 4 input attributes and 1 output attribute or class; two files (numerical and discrete version) for the practical session with 2 input attributes and 1 output attribute or class; and two files (numerical and discrete version) for the online Moodle sessions with 4 input attributes and only one output attribute or class.

Following that, we applied an ensemble or combination of multiple classification base models generated for each of our different sources of data [22]. We used the well-known Vote approach provided by WEKA for automatic combination of machine learning algorithms. This approach tries to combine the probability distributions of these base classifiers. It produces better results than individual classification models if the classifiers of the sets are accurate and diverse. It has demonstrated better results than homogeneous models for standard datasets. Vote adaptively resamples and combines so that resampling weights are increased for those cases more often misclassified and the combination is done by weighted vote. In order to select the best weighting (for each individual classification model) we tested it by giving the same weight (1) or double that (2) to each individual model. The best result with our data was obtained when combining a weight of 1 for Theory and Practical with a weight of 2 for Moodle by using the average as the combination rule for weights.

We executed the six classification algorithms as base or individual classification models of our Voting method for the 6 previously generated summary datasets. Table 4 shows the results (%Accuracy and ROC Area).

**Table 4. Results obtained when using ensembles.**

|  | NUMERICAL DATA | | DISCRETIZED DATA | |
| --- | --- | --- | --- | --- |
|  | % Accuracy | AUC | %Accuracy | AUC |
| **Jrip** | 82.4561 | 0.9230 | **85.9649** | **0.9380** |
| **Nnge** | 77.1930 | 0.8770 | 77.1930 | 0. 8770 |
| **PART** | 80.7018 | 0.9040 | 82.4561 | 0.9130 |
| **J48** | 82.4561 | 0.9110 | 82.4561 | 0.9220 |
| **REPTree** | 82.4561 | 0.9230 | 82.4561 | 0.9220 |
| **Randomtree** | 77.1930 | 0.8360 | 79.9474 | 0.9170 |
| **Avg.** | 80.4093 | 0.8956 | 81.7456 | 0.9185 |

Table 4 shows that the best results (highest values) were produced by Jrip (85.96 %Acc and 0.93 AUC). Once again, on average most of the algorithms exhibited slightly improved performance in both measures when using discretized data.

## 5.4 Experiment 4: Using ensembles and selecting the best attributes

In experiment 4 we applied an ensemble of classification algorithms to the best attributes from each different source of data.

Firstly, we selected the best attributes for each of the three different sets of datasets (6 files in total) generated in experiment 3. For that, we again used the well-known *CfsSubsetEval* attribute selection algorithm, producing the list of attributes shown in Table 5.

**Table 5. Results of attribute selection with CFSSubsetEval.**

| Dataset | Type | # selected features | Name of Selected features |
| --- | --- | --- | --- |
| Theory | Numerical | 2 | Theory.Attendance<br>Theory.Attention |
| | Discretized | 1 | Theory.Attention |
| Practice | Numerical | 2 | Practice.Attendance<br>Practice.Score |
| | Discretized | 2 | Practice.Attendance<br>Practice.Score |
| Moodle | Numerical | 2 | Moodle.Quiz<br>Moodle.Forum |
| | Discretized | 2 | Moodle.Quiz<br>Moodle.Forum |

Following that, we applied an ensemble or combination of multiple classification base models by again using the *Vote* automatic combining machine learning algorithm. To find the best weights (for each individual classification model) we tested it by giving the same weight (1) or double that (2) to each individual model. The best result with our data was obtained when combining a weight of 1 for Theory and Practical with a weight of 2 for Moodle by using the average as combination rule for weights.

We executed the six classification algorithms as base or individual classification models of our Voting method for the 6 previously generated summary datasets. Table 6 shows the obtained results (%Accuracy and ROC Area).

**Table 6. Results obtained when using ensembles and selection of the best attributes.**

| | NUMERICAL DATA | | DISCRETIZED DATA | |
|---|---|---|---|---|
| | % Accuracy | AUC | %Accuracy | AUC |
| **Jrip** | 82.4561 | 0.9170 | 84.2105 | 0.9310 |
| **Nnge** | 80.7018 | 0.9020 | 78.9474 | 0.8900 |
| **PART** | 80.7018 | 0.9010 | 82.4561 | 0.9350 |
| **J48** | 82.4561 | 0.8990 | 84.2105 | 0.9350 |
| **REPTree** | 84.2105 | 0.9130 | **87.4737** | **0.9420** |
| **Randomtree** | 77.1930 | 0.9160 | 82.4561 | 0.9330 |
| **Avg.** | 81.2865 | 0.9080 | 83.2923 | 0.9276 |

Table 6 shows that the best results (highest values) were produced by REPTree (87.47 %Acc and 0.94 AUC). Again, on average, most of the algorithms exhibited slightly improved performance in both measures when using discretized data.

## 5.5 Discussion

Following, we address the two initial research questions by discussing the results produced by our four experiments.

### 5.5.1 Answering question 1

Our first research question was: Which data fusion approach and classification algorithms produce the best results from our data? We used four different data fusion approaches and six white-box classification algorithms to answer this question. The four proposed data fusion approaches were not completely different. They were consecutive, or incremental approaches, each one was a modified or extended version of one or more of the previous approaches:

1. Merging all attributes. Our first data fusion approach which uses a simple approach to Naïve fusion in which general summary statistics are generated by combining the different data sources.

2. Selecting the best attributes. Our second approach (modifying the first approach) in which we applied a reduction of features by selecting the best attributes starting from the previous general summary statistics.

3. Using ensembles. Our third approach (which modified the first approach) applied decision-level fusion to combine the results of 3 classifiers, one for each individual statistical summary of our 3 data sources.

4. Using ensembles and selection of the best attributes. Our fourth approach (Implementation of a hybrid between attribute selection algorithms, classification algorithms and automatic learning algorithms) applied decision-level fusion again combining the results of 3 classifiers but this time having previously selected the best attributes in each of the 3 individual statistical summaries of each data source.

Table 7 shows that the average prediction performance (Average of % Accuracy and AUC) of the classification algorithms increased in each new approach. The second approach improved on the first approach, the third approach improved on the second approach and the best result was produced using the fourth approach of using ensembles and selection of the best attributes. In all the approaches the average values were higher when using discretized data than numerical data.

Table 7. Average results obtained in the four data fusion approaches.

| Average | NUMERICAL DATA | | DISCRETIZED DATA | |
| --- | --- | --- | --- | --- |
| | % Accuracy | AUC | %Accuracy | AUC |
| **Merging all attributes** | 76.2749 | 0.8488 | 77.3567 | 0.8687 |
| **Selecting the best attributes** | 78.9474 | 0.8720 | 80.9942 | 0.8998 |
| **Using ensembles** | 80.4094 | 0.8957 | 81.7456 | 0.9185 |
| **Using ensembles and selection of the best attributes** | 81.2866 | 0.9080 | **83.2924** | **0.9277** |

We were unable to find a single best algorithm that would win in all cases in our experiments (8 cases = 4 experiments * 2 different datasets, numerical and discretized). This is logical and is in line with the No-Free-Lunch theorem [23], in which it is generally accepted that no single supervised learning algorithm can beat another algorithm over all possible learning problems or different datasets. In the first experiment, the algorithm that produced the highest prediction values was PART (80.4561% Accuracy and 0.9170 AUC), in the second experiment it was J48 (82.4561 Accuracy and 0.9230 AUC), in the third it was Jrip (85.9649 Accuracy and 0.9380 AUC), and finally the algorithm that produced the highest prediction values of Accuracy (87.4737%) and AUC (0.9420) was REPTree when using an ensemble and selection of the best attributes from the discretized data in the fourth experiment.

### 5.5.2 Answering question 2

Our second research question was: How useful are the prediction models we produce to help teachers detect students at risk of drop out or fail the course? To answer that, we will demonstrate and describe the prediction model that produced the highest values of Accuracy and AUC in each of our 4 experiments.

In experiment 1, the prediction model that produced the best prediction was generated by the PART algorithm using discretized data (see Table 8).

Table 8. PART decision list when merging all attributes.

| |
| --- |
| IF Moodle.Quiz = High THEN Pass |
| IF Moodle.Quiz = Medium AND Theory.Attention = Medium THEN Pass |
| IF Moodle.Quiz = Low THEN Fail |
| IF Theory.Attention = Low AND Moodle.Forum = Low THEN Dropout |
| ELSE Pass |
| Number of Rules :     5 |

This prediction model (see Table 8) consists of 5 rules that show that the students who had high scores in Moodle quizzes or who had medium scores in Moodle quizzes and also paid attention in theory classes, were the students who passed the course. The students who failed the course were those who got low scores in the Moodle quizzes. The students who dropped out from the course were those who pied little attention in theory classes and also showed low activity in the Moodle forum. The remaining students were classified as passing.

In experiment 2, the prediction model that produced the highest prediction values used the J48 algorithm with the discretized data (see Table 9).

**Table 9. J48 pruned tree when selecting the best attributes.**

```
IF Moodle.Quiz = Low
|  Moodle.Forum = Low
|  |  Theory.Attention = Low THEN Dropout
|  |  Theory.Attention = Medium THEN Fail
|  |  Theory.Attention = High THEN Fail
|  Moodle.Forum = Medium THEN Fail
|  Moodle.Forum = High THEN Fail
ELSE IF Moodle.Quiz = Medium
|  Theory.Attention = Low THEN Fail
|  Theory.Attention = Medium THEN Pass
|  Theory.Attention = High THEN Pass
ELSE IF Moodle.Quiz = High THEN Pass
Number of Leaves:      9
Size of the tree:   13
```

This prediction model (see Table 9) is a decision tree with 9 leaves that can be transformed into 9 prediction rules. These rules show that the students who passed the course are those who had medium scores in Moodle quizzes and also paid medium to high attention in theory classes, or those who simply had high scores in Moodle quizzes. The students who dropped out from the course are those who had low scores in Moodle quizzes, showed low activity in the Moodle forum, and also paid little attention in theory classes. In addition, students who failed are those who had low scores in Moodle quizzes, showed low activity in the Moodle forum and paid medium to high attention in theory classes. There are also other failing student profiles: students who had medium scores in Moodle quizzes and also paid little attention in theory classes; students who had low scores in Moodle quizzes, showed low activity in the Moodle forum, and paid medium to high attention in theory classes.

In experiment 3, the prediction model that produced the highest prediction values used the JRIP algorithm with discretized data (see Table 10).

**Table 10. JRIP when using ensembles.**

| |
|---|
| JRIP rules (Theory): |
| =========== |
| IF (Theory.Attendance = High) THEN Pass |
| IF (Theory.Attention = Low) THEN Dropout |
| ELSE Dropout |
| Number of Rules : 3 |
| |
| JRIP rules (Practice): |
| =========== |
| IF (Practice.Attendance = High) and (Practice.Score = High) THEN Pass |
| IF (Practice.Attendance = Low) and (Practice.Score = Low) THEN Fail |
| ELSE Dropout |
| Number of Rules : 3 |
| |
| JRIP rules (Moodle): |
| =========== |
| IF (Moodle.Task = Low) and (Moodle.Quiz = Low) THEN Fail |
| IF (Moodle.Quiz = Medium) and (Moodle.Forum = Low) THEN Fail |
| IF (Moodle.Task = Medium) THEN Pass |
| IF (Moodle.Quiz = High) THEN Pass |
| ELSE Dropout |
| Number of Rules : 5 |

This prediction model (see Table 10) is a combination of three models that show differential student behavior related to theory, practice and Moodle. The students who regularly attended theory classes passed the course; the students who exhibited low attendance finally dropped out. The students who regularly attended practical classes and exhibited high performance in those practical classes then passed the entire course. In contrast, the students who rarely attended practical classes and had low performance in practicals then failed the entire course. The students who uploaded a moderate number of activities to the Moodle platform or got high scores in Moodle quizzes are students who passed the course; and logically, the students who uploaded a low number of activities to the Moodle platform and got low scores in Moodle quizzes are students who failed the course, but the students with medium performance in quizzes and low contributions to the forum also failed.

In experiment 4, the prediction model that produced the highest prediction values used the RepTree algorithm with discretized data (see Table 11).

**Table 11. RepTree when using ensembles with selecting the best attributes.**

| |
|---|
| REPTree (Theory) |

```
============
IF Theory.Attention = Low THEN Dropout
IF Theory.Attention = Medium THEN Fail
IF Theory.Attention = High THEN Pass
Size of the tree : 4

REPTree (Practice)
============
IF Practice.Attendance = Low THEN Dropout
IF Practice.Attendance = Medium THEN Fail
IF Practice.Attendance = High
|   AND Practice.Score = Low THEN Fail
|   OR Practice.Score = Medium THEN Fail
|   OR Practice.Score = High THEN Pass
Size of the tree : 7

REPTree (Moodle)
============
IF Moodle.Quiz = Low
|   AND Moodle.Forum = Low THEN Dropout
|   OR Moodle.Forum = Medium THEN Fail
|   OR Moodle.Forum = High THEN Fail
ELSE IF Moodle.Quiz = Medium THEN Pass
ELSE IF Moodle.Quiz = High THEN Pass
Size of the tree : 7
```

This prediction model (see Table 11) is also a combination of three models that show differential student behavior related to theory, practicals and Moodle. Exhibiting low attention in theory classes, low practical attendance, or low scores in Moodle quizzes plus little forum participation seems to lead to students dropping out. At the same time, students exhibiting medium or high attention, or medium to high Moodle forum participation, fail; those demonstrating medium practical attendance or high practical attendance plus low or medium practice score also fail. The students that demonstrated high practical attendance and performance passed, as did the students with medium to high scores in Moodle quizzes.

In general, we can see that these white-box models are very useful for explaining to the teacher how the predictions of pass, fail or dropout are arrived at. The teacher can discover what the main predictive attributes and values are directly from the background of the IF-THEN rules. In this sense, the presence of the attributes attention in the classroom, forum participation and score in Moodle quizzes is notable. It is important to notice that all the models produced only had attributes from the theory and online sessions, not from the practical sessions. This may be due to the variables provided/obtained from the practicals were not discriminating in predicting the students' final performance. Our results also revealed that the most discriminant information source was student behavior in Moodle. In this regard, we saw that the two ensemble approaches had optimal weighting when giving greater weight to the on-line data source. Although it is not clear as to how much online learning is inherent in blended learning [24], these results seem to point to the conclusion that the use of distance learning platforms in the b-learning educational experience is positive.

# 6 Conclusions

This paper uses different data fusion approaches in blended learning for answering two research questions:

- Answer to question 1: The use of ensembles and selecting the best attributes approach from discretized summary data produced our highest/best results in Accuracy and AUC values. The REPTree classification algorithm obtained the highest/best results in this approach from discretized summary data.

- Answer to question 2: The white-box models we produced give teachers very understandable explanations (IF-THEN rules) of how they classified the students' final performance or classification. They showed that the attributes that appear most in these rules were attention in theory classes, scores in Moodle quizzes, and the level of activity in the Moodle forum.

As next step, we intend to investigate and do new experiments for trying to improve our process and to overcome some limitations:

- Analyzing the video automatically rather than manually or semi-automatically. Processing the video recordings automatically would gather information more efficiently compared to manual coding [25]. The use of multiple web-cams distributed around the classroom, rather than a single camera, will let us use more advanced algorithms for detecting student engagement more accurately.

- Using raw data and other specific data fusion techniques. We used a basic Naïve and knowledge-based fusion method that uses summary data. However, there are other fusion theories/methods data [13] such as Probability-based methods (PBM) and Evidence reasoning methods (EBM) that we can use with raw data. We could also use semantic (abstract) level features in order to produce intelligent data aggregation.

- Using more sources of information, including videos of practicals and on-line session interaction with Moodle [26], audio from theory classes and practicals, text analytics or text mining of what students write during theory classes, practicals or in Moodle [27]. We can also capture vast amounts of multimodal interaction data by using technologies such as wearable sensors and biosensors that measure skin conductivity, heartbeat, electroencephalography, gesture sensing, eye tracking, etc. And, of course, the new source of information included in the latest version of Moodle, a Learning Analytics API that would let us explore the data stored in Moodle from new perspectives.

- Making data representation more human understandable: It would be advisable to think in higher-level variables to facilitate the analysis of student behavior and produce sounder models which were easier to interpret. These variables could be generated either by the teachers or educational science experts, and could be in line with fuzzy labels such as engagement, motivation, procrastination, frustration, etc.

- Broadening and testing the potential applicability to other educational settings that are going to become essential in the current and near future educational context such as Personal Learning Environments (PLEs), Intelligent Tutoring Systems (ITS), Hypermedia learning Environments, Massive Open Online Courses (MOOCs), etc. In short, make our approach transferrable to any blended Computer Based Learning Environment that gives support to the current teaching-learning process mediated by the current pandemic-related conditions.

**ACKNOWLEDGMENTS**

This work would not be possible without the funding from the Ministry of Sciences and Innovation I+D+I (TIN2017-83445-P and PID2019-107201GB-100) and the European Regional Development Fund and the Principality of Asturias (FC-GRUPIN-IDI/2018/000199).